\documentclass[twocolumn,showpacs,preprintnumbers,amsmath,amssymb,APSl,prd,nofootinbib,superscriptaddress]{revtex4-1}
\usepackage{dcolumn}
\usepackage{bm}
\usepackage{ifpdf}
\usepackage{hyperref}
\usepackage{dcolumn}
\usepackage{bm}
\usepackage[spanish,english]{babel}
\usepackage{amsfonts}
\usepackage{amssymb}
\usepackage{graphicx}
\usepackage[latin1]{inputenc}

\newcommand{\be}{\begin{equation}}
\newcommand{\en}{\end{equation}}
\newcommand{\bea}{\begin{eqnarray}}
\newcommand{\ena}{\end{eqnarray}}
\begin{document}
\title{Thick brane in $f(R)$ gravity with Palatini dynamics}
\author{D. Bazeia} \email{bazeia@fisica.ufpb.br}
\affiliation{Departamento de F\'isica, Universidade Federal da
Para\'\i ba, 58051-900 Jo\~ao Pessoa, Para\'\i ba, Brazil}
\author{L. Losano} \email{losano@fisica.ufpb.br}
\affiliation{Departamento de F\'isica, Universidade Federal da
Para\'\i ba, 58051-900 Jo\~ao Pessoa, Para\'\i ba, Brazil}
\author{R. Menezes}\email{rmenezes@dce.ufpb.br}
\affiliation{Departamento de Ci\^encias Exatas, Universidade Federal da
Para\'iba, 58297-000 Rio Tinto, PB, Brazil}
\affiliation{Departamento de F\'\i sica, Universidade Federal de Campina
Grande, 58109-970 Campina Grande, PB, Brazil}
\author{Gonzalo J. Olmo} \email{gonzalo.olmo@csic.es}
\affiliation{Departamento de F\'{i}sica Te\'{o}rica and IFIC, Centro Mixto Universidad de Valencia - CSIC.
Universidad de Valencia, Burjassot-46100, Valencia, Spain}
\affiliation{Departamento de F\'isica, Universidade Federal da
Para\'\i ba, 58051-900 Jo\~ao Pessoa, Para\'\i ba, Brazil}
\author{D. Rubiera-Garcia} \email{drgarcia@fc.ul.pt}
\affiliation{Instituto de Astrof\'isica e Ci\^encias do Espa\c{c}o, Universidade de Lisboa, Faculdade de Ci\^encias, Campo Grande, PT1749-016 Lisboa, Portugal}
\affiliation{Center for Field Theory and Particle Physics and Department of Physics, Fudan University, 220 Handan Road, 200433 Shanghai, China}
\affiliation{Departamento de F\'isica, Universidade Federal da
Para\'\i ba, 58051-900 Jo\~ao Pessoa, Para\'\i ba, Brazil}
\begin{abstract}
This work deals with modified gravity in five dimensional spacetime. We study a thick Palatini $f(R)$ brane, that is, a braneworld scenario described by an anti-de Sitter warped geometry with a single extra dimension of infinite extent, sourced by real scalar field under the Palatini approach, where the metric and the connection are regarded as independent degrees of freedom. We consider a first-order framework which we use to provide exact solutions for the scalar field and warp factor. We also investigate a perturbative scenario such that the Palatini approach is implemented through a Lagrangian $f(R)=R+\epsilon R^n$, where the small parameter $\epsilon$ controls the deviation from the standard thick brane case. In both cases it is found that that the warp factor tends to localize the extra dimension due to the non-linear corrections.
\end{abstract}
\maketitle

\section{Introduction}
\label{sec:intro}

Investigations dealing with spacetime engendering higher spatial dimensions started in physics soon after the appearance of General Relativity (GR) through the Kaluza-Klein models (see e.g. \cite{KK} for a review), aimed to study unification of the electromagnetic interaction with gravity. Nowadays, the presence of higher spatial dimensions is very natural in high energy physics, in string, superstring and other unification and fundamental theories \cite{strings}. However, the addition of extra spatial dimensions is in conflict with the natural world which, when probed in any experiment, has only revealed the presence of three spatial dimensions, though proposals using large extra dimensions with potential experimental signatures in particle accelerators have been discussed \cite{large}.

To reconcile the constraint of three spatial dimensions of the natural world with the introduction of extra dimensions, important scenarios have been proposed. Here we focus our attention upon the Randall-Sundrum (RS) work \cite{rs2}, where the relevant portion of the higher dimensional spacetime is embedded within a five-dimensional anti-de Sitter $(AdS_5)$ geometry. This scenario assumes that the $(3,1)$ spacetime that describes the natural world is embedded in an $AdS_5$ warped geometry, with a single extra spatial dimension of infinite extent. This is known as the RS2 braneworld scenario, and the warp factor identifies a thin brane profile, decaying along the extra dimension $y$ in the form $\exp(-2|y|)$.

Soon after the proposed thin braneworld scenario, it was modified with the presence of scalar fields, giving rise to a new, very interesting thick braneworld scenario, in which the warp factor is now described by another function, which depends on the specific scalar field model one considers \cite{gw,f}. The presence of scalar fields brought interesting possibilities, as the appearance of a new feature, the splitting of the brane, which springs under the presence of distinct effects, at finite temperature \cite{c} or with specific scalar field models \cite{bfg} in the presence of at least one additional parameter, to be used to control the splitting of the brane.

Like the number of spatial dimensions, there are other foundational aspects of the idea of gravitation as a geometric phenomenon which are likely to provide interesting new viewpoints on the fundamental open questions of gravitational physics. In this sense, most extensions of GR adopt the implicit assumption that spacetime is a Riemannian structure completely determined by the metric degrees of freedom (see e.g. \cite{review-MG} for some reviews). However, though geometry made its appearance in physics through Einstein's theory of gravity, there exist many other physical systems in which geometry plays an important role. For example, ordered structures such as Bravais crystals (see e.g. \cite{Kittel}), graphene, and other solid state systems admit a continuum description in terms of differential geometry \cite{Kroner1}. It turns out that while idealized crystals without defects can be described in terms of Riemannian geometry, all known ordered structures, which  present defects of different kinds, require a non Riemannian description \cite{Kroner2} involving non-metricity  \cite{Falk81} and torsion \cite{Kroner3} among other geometrical structures. This fact suggests that 
metric-affine geometry could be favored in nature and its implications for gravitational physics should be explored in detail \cite{lor-c}. This observation is of fundamental importance, since it puts forward that the nature of the underlying space-time geometry is a question that must be determined by observation rather than imposed by convention or selected on practical grounds. For this reason, one can legitimately consider new geometric scenarios and explore their phenomenology to gain insight on the kind of new physics that they could bring about. In this sense, the notion of braneworld scenario is similar in some respects to that of interfaces or thin films within crystals. The geometric elements required for a consistent description of such condensed matter systems should be incorporated in gravitational constructions of the braneworld type and their physical implications scrutinized. This might bring useful new viewpoints on the issue of higher-dimensional models of the physical world.

In the last years, some of us have carried out a program where black hole solutions have been obtained within extensions of GR formulated \`{a} la Palatini, i.e., assuming that metric and connection are independent geometrical entities. Though in GR metric and Palatini formulations lead to the same field equations, this ceases to be the case as soon as one considers extensions beyond GR. New gravitational physics can thus be found in four-dimensional theories formulated in the Palatini formalism, like $f(R)$ \cite{or1}, $f(R,R_{\mu\nu}R^{\mu\nu})$ \cite{or2}, Born-Infeld gravity \cite{or3} and also in five-dimensional $f(R)$ gravity \cite{blor}, which differs from the usual metric formulation of those same theories. A nice feature of these Palatini theories, in particular, is that the point-like black hole singularity found in GR is generically replaced by a finite area wormhole structure, with potentially relevant consequences for the understanding of the last stages of black hole evaporation \cite{lor}, the information loss problem \cite{Chen:2014jwq},
and the phenomenology of black holes at particle accelerators \cite{OR14}.

The aim of this work is to offer a first step to combine the two fundamental questions on the foundations of space-time raised above. We explore the theoretical and phenomenological implications for braneworld scenarios of admitting that the spacetime has independent metric and affine structures, as in the Palatini approach \cite{W,pala1}. We shall thus focus on the thick braneworld scenario with a single extra dimension of infinite extent, considering the presence of a real scalar field in the $AdS_5$ geometry and assuming the gravitational dynamics to be described by a Palatini $f(R)$ Lagrangian (see \cite{pala1,pala2} for reviews), as the simplest extension of GR. See also Ref.~\cite{FRB} for previous studies on thick braneworld scenarios with $f(R)$ dyamics in the metric approach. We note that in the Palatini formulation, the vacuum field equations exactly boil down to the equations of GR with, possibly, a cosmological constant (depending on the particular gravity Lagrangian chosen). Nevertheless, when matter fields are present, modified dynamics arises. Therefore, Palatini theories offer a way to generate new gravitational effects without the need for introducing new dynamical degrees of freedom.

We start in Sec.~\ref{sec:gen} by introducing notation and the model to be investigated, which describes a source scalar field minimally coupled to the Palatini $f(R)$ geometry. We then deal with the braneworld scenario in Sec.~\ref{sec:brane}, and there we write the equations of motion and the scalar field equations in a first-order framework. Also, we verify consistency of the equations of motion that appear in the Palatini braneworld scenario, which further reduce to the equations of motion of the standard thick braneworld scenario when one changes $f(R)\to R$, as expected. We next consider two different approaches to solve the first-order equations. First, we study a simple example of a Palatini brane in Sec.~\ref{sec:example}, where one begins with a specific gauge function and reconstructs the scalar field and the $f(R)$ model in an implicit manner, finding a new interesting effect, namely, that the warp factor vanishes asymptotically, much faster than it does in the case of a standard thick brane. In the second approach, in Sec.~\ref{sec:papproach} we consider a small correction to the GR Lagrangian, $f(R)=R+\epsilon R^n$, with $n=2,3,...,$ where $\epsilon$ is a small parameter. We then implement a perturbative procedure, obtaining results valid up to first-order in $\epsilon$. We study two distinct examples, one with the potential of the source scalar field being polynomial, up to the $\phi^4$ power and engendering spontaneous symmetry breaking, and the other nonpolynomial, of the sine-Gordon type. The results are compared with the cases of standard thick branes, with $\epsilon=0$, and they appear to behave consistently, suggesting that the new braneworld scenarios are robust. Thus, the generic first-order framework allows us to study both exact new brane configurations and perturbative departures from the GR dynamics. In Sec.~\ref{sec:end} we end the work with some comments and conclusions.

\section{Field equations for five-dimensional Palatini $f(R)$ gravity}
\label{sec:gen}

We start with the action
\be \label{eq:action}
S=\frac{1}{2\kappa^2} \int d^5x \sqrt{-g} f(R)  + {S_s}(g_{\mu\nu},\varphi),
\en
where $\kappa^2$ is Newton's gravitational constant in appropriate system of units (in GR, $\kappa^2 =8\pi G$), $g$ is the determinant of the spacetime metric $g_{\mu\nu}$, $R=g^{\mu\nu}R_{\mu\nu}(\Gamma)$ is the curvature scalar constructed with the Ricci tensor $R_{\mu\nu}(\Gamma)=\partial_\lambda \Gamma^\lambda_{\nu\mu}-\partial_\nu\Gamma^\lambda_{\lambda \mu}+\Gamma^\lambda_{\lambda \kappa}\Gamma^\kappa_{\nu \mu}-\Gamma^\lambda_{\nu\kappa}\Gamma^\kappa_{\lambda \mu}$, where the connection $\Gamma \equiv \Gamma_{\mu\nu}^{\lambda}$ is a priori independent of the metric (Palatini formalism). For simplicity we assume a torsionless scenario,  $\Gamma_{\mu\nu}^{\lambda}=\Gamma_{\nu\mu}^{\lambda}$ (see \cite{Olmo:2013lta} for more details on the role of torsion in Palatini theories). The source contribution ${S_s}$ is supposed to couple to the metric only, and $\varphi$ denotes collectively the source fields. The spacetime is five-dimensional, so $\mu,\nu,...=0,1,...,4;$ also, we will use latin indices $a,b,...=0,1,2,3$ to span the four-dimensional spacetime, and denote the fifth dimension $x^4$ by $y$.

The field equations for the action (\ref{eq:action}) are obtained by independent variation with respect to the metric and connection. The details of this derivation can be found in \cite{blor} and therefore we bring here the final results:
\bea
f_R R_{\mu\nu} - \frac{1}{2} f g_{\mu\nu}&=&\kappa^2 T_{\mu\nu}, \label{eq:metric} \\
\nabla^{\Gamma}_{\lambda} (\sqrt{-g} f_R g^{\mu\nu})&=&0, \label{eq:connection}
\ena
where we have defined $f_R \equiv df/dR$, while $T_{\mu\nu}=-\frac{2}{\sqrt{-g}} \frac{\delta S_s}{\delta g^{\mu\nu}}$ is the energy-momentum tensor of the matter, and $\nabla^{\Gamma}_{\lambda}$ denotes the covariant derivative with respect to the independent connection $\Gamma$. We emphasize that the field equations above are different from those corresponding to the usual metric (or Riemannian) formulation of $f(R)$ theories. In fact, the metric formulation leads to higher-order equations for the metric, whereas the Palatini formulation presented here yields second-order equations, as will be shown next.

To solve the above dynamical equations, we introduce an auxiliary metric $h_{\mu\nu}$ so that Eq.~(\ref{eq:connection}) can be formally written as
\be
\nabla^{\Gamma}_{\lambda}(\sqrt{-h} h^{\mu\nu})=0 \label{eq:hG}
\en
Comparison between (\ref{eq:hG}) and (\ref{eq:connection}) leads to
\be \label{eq:g-h}
h_{\mu\nu}=f_R^{2/3} g_{\mu\nu};\;\;\; h^{\mu\nu}=f_R^{-2/3} g^{\mu\nu},
\en
which puts forward that the two metrics are conformally related. We note that these equations imply that the independent connection $\Gamma$ is metric-compatible with $h_{\mu\nu}$ (but {\it not} with $g_{\mu\nu}$), which implies that $\Gamma_{\mu\nu}^{\lambda}$ is the Levi-Civita connection of $h_{\mu\nu}$ (see \cite{W} for details). The introduction of this auxiliary metric greatly simplifies the explicit expression of the metric field equations (\ref{eq:metric}), which read \cite{blor}
\be \label{eq:fieldequations}
{R_\mu}^{\nu}(h)=\frac{\kappa^2}{f_R^{5/3}} \left( \frac{f}{2\kappa^2}{\delta_\mu}^\nu +{T_\mu}^{\nu} \right),
\en
where ${T_\mu}^{\nu}=T_{\mu\alpha}g^{\alpha\nu}$.
We note from Eq.~(\ref{eq:g-h}) that for $f(R)=R$, the new metric $h_{\mu\nu}$ coincides with $g_{\mu\nu}$, and we then get back to GR. It is worth mentioning i) the second-order character of the field equations (\ref{eq:fieldequations}) and ii) the fact that in vacuum, ${T_\mu}^{\nu}=0$, they boil down to the equations of GR plus a cosmological constant term (depending on the explicit functional form of the $f(R)$ Lagrangian chosen), which implies absence of new propagating degrees of freedom in the spectrum of the theory. It is also worth noting that if the field equations (\ref{eq:fieldequations}) are written in terms of $g_{\mu\nu}$ instead of $h_{\mu\nu}$ (which is the form we use in this work), then one finds an Einstein-like set of equations of the form $G_{\mu\nu}(g)=\tau_{\mu\nu}$, where $G_{\mu\nu}(g)$ represents the Einstein tensor and  $\tau_{\mu\nu}$ is an effective stress energy tensor (see the review \cite{pala1} for details). Written in that form, it is easy to verify that the Bianchi identity $\nabla_\mu {G^\mu}_{\nu} (g)=0$ is trivially satisfied, while the conservation of $\tau_{\mu\nu}$ requires the use of the matter field equations.

Further information on this problem can be obtained by taking the trace in Eq.~(\ref{eq:metric}), which yields
\be\label{eq:trace}
Rf_R - \frac{5}{2} f=\kappa^2 T \ .
\en
We note that this is not a differential equation, but just an algebraic nonlinear relation between the gravity and source field functions defining the model, and extending the five-dimensional GR relation $R=-\frac{2}{3} \kappa^2 T$ to the nonlinear $f(R)$ case, where $R=R(T)$ is a non-linear function of the trace $T$. As a result, all the functions of $R$ appearing in the field equations must be interpreted as functions of the matter fields via $R=R(T)$. This fact has two important consequences: i) the conformal factor relating the metrics $h_{\mu\nu}$ and $g_{\mu\nu}$ is determined by the matter fields and does not require solving an independent dynamical equation, and ii) the energy-density of the matter fields can be seen as playing a role analogous to that of the density of point defects in a hypothetical space-time microstructure \cite{lor-c}.

Let us point out that there is a well known equivalence between $f(R)$ theories in Palatini formalism and a particular case of Brans-Dicke scalar-tensor theory \cite{Olmo2005}. In this alternative representation, the scalar field turns out to be non-dynamical, as it simply encodes in a different language the nonlinear relation between matter and curvature that typically arises in the Palatini approach. Since the scalar-tensor representation does not bring any useful new insight or simplification in our analysis, in this work we prefer not to use it.

\section{Braneworld setup}
\label{sec:brane}

Inspired by the braneworld scenarios described in Refs.~\cite{rs2,gw,f,c,bfg}, we consider the following source action, using signature $(-,++++)$
\be
S_s=\frac{1}{2} \int d^5x \sqrt{-g} \left( g^{\mu\nu} \partial_{\mu}\phi \partial_{\nu} \phi + 2V(\phi)\right),
\en
with $\phi$ a real scalar field and $V(\phi)$ the corresponding potential. The energy-momentum tensor for this source field follows as
\be
T_{\mu\nu}=\partial_{\mu}\phi \partial_{\nu}\phi-\frac{1}{2} g_{\mu\nu}(g^{\alpha\beta}\partial_{\alpha} \phi \partial_{\beta} \phi + 2V(\phi)).
\en
We are interested in the spacetime metric
\be
ds^2=e^{2A(y)} g_{ab} dx^a dx^b + dy^2,
\en
where $\exp(2A(y))$ is the warp factor, which is assumed to depend only on the extra dimension. This line element describes an $AdS_5$ warped geometry with an extra spatial dimension of infinite extent, similar to the standard RS2 braneworld scenario. Using the relation between $h_{\mu\nu}$ and $g_{\mu\nu}$ given by (\ref{eq:g-h}) we can write the line element for $h_{\mu\nu}$ as
\be \label{eq:line}
d\tilde{s}^2=h_{\mu\nu} dx^\mu dx^\nu=f_R^{2/3} e^{2A(y)} g_{ab}dx^a dx^b +f_{R}^{2/3} dy^2.
\en
As it appears in the standard braneworld scenario, here we also assume that the scalar field only depends on the extra dimension. Thus, its energy-momentum tensor reads
\be \label{eq:Tmunu}
{T_\mu}^{\nu}=
\left(
\begin{array}{cc}
-\frac{1}{2} (\phi_y^2+2V)  I_{4 \times 4} &  \hat{0} \\
\hat{0}  & \frac{1}{2} (\phi_y^2 - 2V)  \\
\end{array}
\right),
\en
where $\phi_y \equiv d\phi/dy$ and $I_{4 \times 4}$ is the four-dimensional identity matrix. This allows to easily obtain the trace as $T=-5V-\frac{3}{2} \phi_y^2$. Putting these elements into the field equations (\ref{eq:fieldequations}) we get
\be \label{eq:gravity}
{R_\mu}^{\nu}\!(h)\!=\!\frac{1}{2f_R^{5/3}}\!\!
\left(
\begin{array}{cc}
\!\!\!(f\!-\!\kappa^2 (\phi_y^2\!+2V))  I_{4 \times 4}\!\!\! &  \hat{0} \\
\hat{0}  & \!f\! + \kappa^2 (\phi_y^2\! - 2V) \!\!\! \\
\end{array}
\right).
\en
On the other hand, the source scalar field equation reads
\be \label{eq:matter}
\phi_{yy}+4A_y \phi_y=V_{\phi}.
\en
with $A_y \equiv dA/dy$. We note that the system of equations with the symmetries of the problem gives rise to three equations to be solved, though they are not fully independent and one of them can be deduced from the other two, a result that can be traced back to the existence of Bianchi identities in our (torsionless) scenario, as pointed out above. This gives consistency to the problem since we have two independent variables, $A(y)$ and $\phi(y)$. This issue also appears in the standard braneworld scenario.

Using the fact that the coefficients of the independent connection $\Gamma$ correspond to the Christoffel symbols of the metric $h_{\mu\nu}$ (recall Eq.(\ref{eq:hG})) we obtain, after standard calculations, the components of the Ricci tensor
\bea
{R_i}^i(h)&=&-\frac{1}{3f_R^{2/3}} \left(7A_y \frac{f_{R,y}}{f_R} \!+\! 3(4A_y^2\! +\!A_{yy})\!+\!\frac{f_{R,yy}}{f_R} \right)  \\
{R_4}^4(h)\!&=&\! - \frac{4}{3f_R^{2/3}}\! \!\left(\!-\frac{f_{R,y}^2}{f_R^2} \!+\! 3(A_y^2\! +\!A_{yy} )\!+\! A_y \frac{f_{R,y}}{f_R}
\! + \!\frac{f_{R,yy}}{f_R} \! \right). \nonumber
\ena
We now manipulate these equations to put them in a more amenable form. Consider the combination $4{R_i}^i-{R_4}^4$, which yields
\be \label{eq:field1}
\left(A_y + \frac{1}{3} \frac{f_{R,y}}{f_R} \right)^2=-\frac{1}{8f_R} \left(f-\kappa^2 \left( 2V+\frac{5}{3} \phi_y^2 \right) \right),
\en
where the object $f_{R,y}/f_R$ can be computed as
\be \label{eq:fieldauxiliary}
\frac{f_{R,y}}{f_R}=\frac{2\kappa^2 f_{RR}}{f_R (Rf_{RR}-\frac{3}{2} f_R)} ( 6A_y \phi_y-V_\phi)\phi_y.
\en
Note that in the GR limit, $f_R=1$, Eq.~(\ref{eq:field1}) nicely recovers the right expression, $A_y^2=\frac{\kappa^2}{12} (\phi_y^2 + 2V)$ (with $\kappa^2=2$). See, e.g., Refs.~\cite{f}.

On the other hand, the combination ${R_i}^i-{R_4}^4$ yields
\bea \label{eq:field2}
3\Big(A_{yy} &+& \frac{1}{3} \frac{f_{R,yy}}{f_R}-\frac{1}{3} \Big(\frac{f_{R,y}}{f_R} \Big)^2\Big) \nonumber \\ &=&
 \frac{f_{R,y}}{f_R} \Big(A_y+ \frac{1}{3} \frac{f_{R,y}}{f_R} \Big) - \frac{\kappa^2}{f_R}\phi_y^2.
\ena
We then define
\be \label{eq:Qdefinition}
\theta \equiv A_y + \frac{1}{3} \frac{f_{R,y}}{f_R},
\en
and using the trace equation, we can turn (\ref{eq:field1}) into
\be \label{eq:field3}
\theta^2=\frac{2\kappa^2 \phi_y^2}{15f_R} - \frac{R}{20},
\en
On the other hand,  (\ref{eq:field2}) can be rewritten as
\be \label{eq:field4}
\theta_y=\frac{f_{R,y}}{3f_R}\theta - \frac{\kappa^2 \phi_y^2}{3f_R}.
\en

\subsection{Consistency of the model}

So far we have obtained the metric field equations (\ref{eq:field3}) and (\ref{eq:field4}). Conservation of energy-momentum, which follows from the metric field equations, must imply on consistency grounds the scalar field equation (\ref{eq:matter}). We now proceed to verify this point. We begin by taking a derivative of  (\ref{eq:field3}) with respect to $y$ and then using (\ref{eq:field4}) to replace $\theta_y$. This leads to
\begin{equation}\label{eq:step_1}
\frac{2}{3}\frac{f_{R,\phi}}{f_R}\theta^2-\frac{2\kappa^2}{3f_R}\left(\theta-\frac{1}{5}\frac{f_{R,\phi}}{f_R}\phi_y\right)\phi_y+\frac{R_\phi}{20}=\frac{4\kappa^2}{15f_R}\phi_{yy} \ .
\end{equation}
Replacing $\theta^2$ in this equation with (\ref{eq:field3}) and $\theta$ with \ref{eq:Qdefinition}, we get
\begin{equation}\label{eq:step_2}
A_y\phi_y+\frac{1}{20\kappa^2}\left(R f_{R,\phi}-\frac{3}{2}R_\phi f_R\right)=-\frac{2}{5}\phi_{yy} \ .
\end{equation}
Here is where we can use the scalar field equation. The trick is to use $A_y\phi_y=-(\phi_{yy}-V_\phi)/4$, which leads to
\begin{equation}
5\kappa^2 V_\phi+3\kappa^2 \phi_{yy}=-\left(R f_{R,\phi}-\frac{3}{2}R_\phi f _R\right)\ .
\end{equation}
Given that $\phi_{yy}=\frac{1}{2}\frac{d \phi_y^2}{d\phi}$ and that $R f_{R,\phi}-\frac{3}{2}R_\phi f _R=\frac{d}{d\phi}\left(R f_R-\frac{5}{2}f\right)$, we can rewrite (\ref{eq:step_2}) in the form
\begin{equation}\label{eq:step_3}
-\kappa^2\frac{d}{d\phi}(5 V+\frac{3}{2}\phi_y^2)=\frac{d}{d\phi}\left(R f_R-\frac{5}{2}f\right) \ ,
\end{equation}
which is just the derivative with respect to $\phi$ of the trace equation (\ref{eq:trace}) with the matter described by (\ref{eq:Tmunu}), for which $T=-\left(5V+\frac{3}{2}\phi_y^2\right)$.
This verifies the consistency of the metric and scalar field equations.

\subsection{First-order equations}

We now go further into the above Palatini thick brane scenario and, in analogy with the standard GR problem, we search for first-order differential equations able to solve the equations of motion. In order to extend the procedure of Ref.~\cite{foe}  to nonlinear $f(R)$ Lagrangians, we introduce two new functions $W=W(\phi)$ and $\alpha=\alpha(\phi)$, and we write $\theta$ as
\begin{equation}
\theta=-\frac{1}{3}\alpha(\phi) W(\phi) \ ,
\end{equation}
with $\alpha(\phi)$ an unspecified function of $\phi$. Computing $\theta_y$ and inserting the result in (\ref{eq:field4}), we get
\be \label{eq:step1}
W\left(\frac{\alpha}{3}\frac{f_{R,\phi}}{f_R} -\alpha_\phi\right)+\frac{\kappa^2\phi_y}{f_R}=\alpha W_\phi \ .
\en
We note that for $f=R$, we get from (\ref{eq:Qdefinition}) that $\theta=A_y$, and the above equation becomes
\be \label{eq:phiy}
\phi_y=\frac12\,W_\phi,
\en
with the choices $\kappa^2=2$ and $\alpha=1$. This is the expected result, which recovers the standard braneworld scenario of GR. Inspired on this, we now make a different choice, considering $\alpha=f_R^{1/3}$. In this case, we get a very natural extension of the standard results, using
\be \label{eq:supertheta}
\theta= -\frac{1}{3}f_R^{1/3} W .
\en
The first-order equations are
\begin{eqnarray}
\phi_y&=&\frac{f_R^{4/3}}{\kappa^2} W_\phi \label{eq:superphi} \\
A_y&=& -\frac{1}{3}f_R^{1/3} \left(W+\frac{1}{\kappa^2}f_{R,\phi}W_\phi\right) \label{eq:Ay}\label{eq:Ay}\,.
\end{eqnarray}
which represents the counterpart of the GR results, valid for any $f(R)$ gravity model in Palatini approach. The simplicity of these equations is in sharp contrast with the strategy followed in Ref.\cite{Gu} to obtain analytic solutions in a similar scenario by ``brute force".

To go further, we insert the results (\ref{eq:supertheta}) and (\ref{eq:superphi}) into  (\ref{eq:field3}) to get
\begin{equation}
\frac{R}{20}=\frac{f_R^{2/3}}{3}\left(\frac{2}{5\kappa^2}f_R W_\phi^2-\frac{1}{3}W^2\right) \label{eq:R/20}
\end{equation}
We note now that by specifying the functions $W(\phi)$ and $f_R(\phi)$, one automatically gets $R(\phi)$ from (\ref{eq:R/20}). It is then immediate to obtain $f(\phi)=\int f_R(\phi) dR(\phi)$. This yields a parametric representation for $f[R(\phi)]$.

With the above results it is easy to verify that for $f=R$, $\alpha=1$, and $\kappa^2=2$ the results of GR \cite{foe} are nicely recovered. In fact,  from (\ref{eq:superphi}) one obtains (\ref{eq:phiy}), and from (\ref{eq:Ay}) the first-order equation
\be \label{foe2}
A_y=-\frac13\, W.
\en
Also, we make simple algebraic manipulations to write
\be \label{foe3}
V(\phi)=\frac18\, W_\phi^2-\frac13\,W^2.
\en
These results lead  to the correct first-order Eqs.~\ref{eq:phiy} and \ref{foe2}, with the potential \ref{foe3}, as they appear in the standard GR scenario.

\section{Example}
\label{sec:example}

Let us now work out a simple example of a Palatini brane. We first use (\ref{eq:Ay}) to have the equation for the warp function in the form
\be\label{wfunction}
A(\phi)=-\frac13 \ln f_R -\frac{\kappa^2}{3} \int d\phi f^{-1}_R \frac{W}{W_\phi} - A_0
\en
where $A_0=A(0)$.  A non-zero value for this constant could lead to asymmetric thick brane scenarios where the geometries on either side of the brane become different. Now, we suppose that the scalar field obeys
\be \label{eq:phi}
\phi_y=b\cos(b\phi),
\en
where $b$ is real parameter, whose value determines both the width of the domain wall and the warp factor [see Eq.(\ref{warpfunctionSOL}) below]. Note that  the limit $b \rightarrow \infty$ yields a singular domain wall (D3-brane) and thus in what follows $b$ will be assumed to be finite. Eq.(\ref{eq:phi}) can be integrated to give the solution
\be
\phi(y)=\frac{1}{b}\arcsin(\tanh(b^2y))\,.
\en
Also, we suppose that
\be
f_R=(1+ab^2\cos^2(b\phi))^{-3/4}\,,
\en
where $a$ is another real parameter, which controls the deviations from GR solutions [recovered for $a \rightarrow 0$]. Thus, from Eq.~\ref{eq:superphi} we get
\be
W(\phi)=\kappa^2\sin(b\phi)\left(\left(1+\frac{2ab^2}{3}\right)+ \frac{ab^2}{3}\cos^2(b\phi)\right)\,.
\en

Thus, the warp function \ref{wfunction} becomes
\begin{eqnarray}\label{warpfunctionSOL}
A(y)&=& -A_0+\ln(u) +\frac{2\kappa^2}{27b^2}u^3-\\
&&\frac{\kappa^2}{3b^2}\left(1+\frac{2ab^2}{3}\right)\bigl({\rm arctanh}(1/u)+\arctan(u)\bigr)\nonumber\,,
\end{eqnarray}
where $u(y)=(1+ab^2 S(y))^{1/4}$ and $S(y)={\rm sech}^2(b^2y)$.
Moreover, the scalar curvature can be written as
\begin{eqnarray}\label{curvatureRY}
R(y)&=&\frac{20}{3}u^4\,\Biggl({\frac {2b^2}{5}}\,{S}^{2}u^5\!-\!\frac13 \kappa^2\biggl(\left(1+\frac{2ab^2}{3}\right)-
\nonumber\\
&&\left(1+\frac{ab^2}{3}\right)S^2-\frac{ab^2}{3} S^4 \biggr)\Biggr).
\end{eqnarray}

\begin{figure}[t]
\includegraphics[{height=3.2cm,angle=-00}]{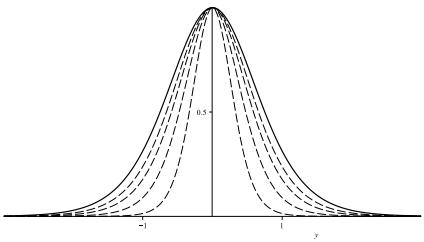}
\caption{Profile of the warp function $e^{2A(y)}$, where $A(y)$ is given by Eq.~\ref{warpfunctionSOL}. We take $b=1$, $\kappa^2=2$, and $a=0$ (solid line), $0.125,0.25,0.5$ and $1.0$ (dashed lines).} \label{fig:1}
\end{figure}
\begin{figure}[t]
\includegraphics[{height=3.2cm,angle=-00}]{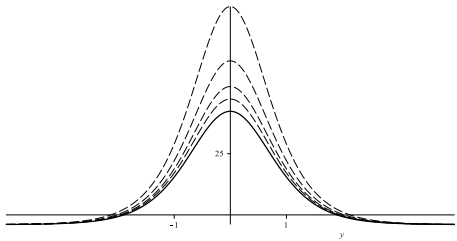}
\caption{Profile of the curvature $R(y)$ given by Eq.~\ref{curvatureRY}. We take $b=1$, $k^2=2$, and $a=0$ (solid line)$,0.125,0.25,0.5$ and $1.0$ (dashed lines).}\label{fig:2}
\end{figure}


We depict the warp factor \ref{warpfunctionSOL} in Fig.~\ref{fig:1}, and the scalar curvature \ref{curvatureRY} in Fig.~\ref{fig:2}.
We note from the warp factor, that for $b=1$ and $\kappa=2$, the value $a=0$ reproduces the GR result, and the increasing of the parameter $a$ tends to localize further the extra dimension. This is an interesting effect, not found in the standard thick brane scenario.

\section{Perturbative approach}
\label{sec:papproach}

In order to further probe the Palatini brane scenario, in this section we study models that can be seen as small fluctuations around the standard thick brane scenario. We take units $\kappa^2=2$ and investigate Palatini brane, governed by a real, very small parameter $\epsilon$, which in the limit $\epsilon\to0$ leads us back to the case of a standard thick brane. This possibility is implemented using
\begin{equation}
f(R)=R+\epsilon\,R^n\,,
\end{equation}
with $n$ integer, $n=2,3,...,$. This model, with $n=2$, was considered in \cite{or1} to study charged black hole (Reissner-Nordstr\"om-like) solutions and also to study inflation (see e.g. \cite{inflation}) and non-singular cosmologies \cite{Barragan:2009sq}. Here, this example will allow us to implement a standard perturbative approach, going up to first order in
$\epsilon$.

With this in mind, we use \ref{eq:R/20} to obtain
\begin{equation}
R=R_0(\phi)-\epsilon\, c(n) \,R_0^{n-1}(\phi)\,W^2\,W_\phi^2\,,
\end{equation}
where
\begin{equation}
R_0(\phi)=\frac{4}{3}W_\phi^2-\frac{20}{9}W^2\,,\,\,\,c(n)=\frac{7n}{405}\,.
\end{equation}
The potential takes the form
\begin{equation}\label{vaprox}
V(\phi)=\frac18W_\phi^2-\frac13W^2 -\epsilon V_\epsilon(\phi)\,,
\end{equation}
with the contribution $V_\epsilon$ given by
\begin{equation}
V_\epsilon(\phi)=\frac1{20}\left((2n-5)R_0+\left(4+3\,c(n)W^2\right)W_\phi^2 R_0^{n-1}\right)\,.
\end{equation}

\begin{figure}[t]
\includegraphics[{height=3.0cm,width=8cm,angle=-00}]{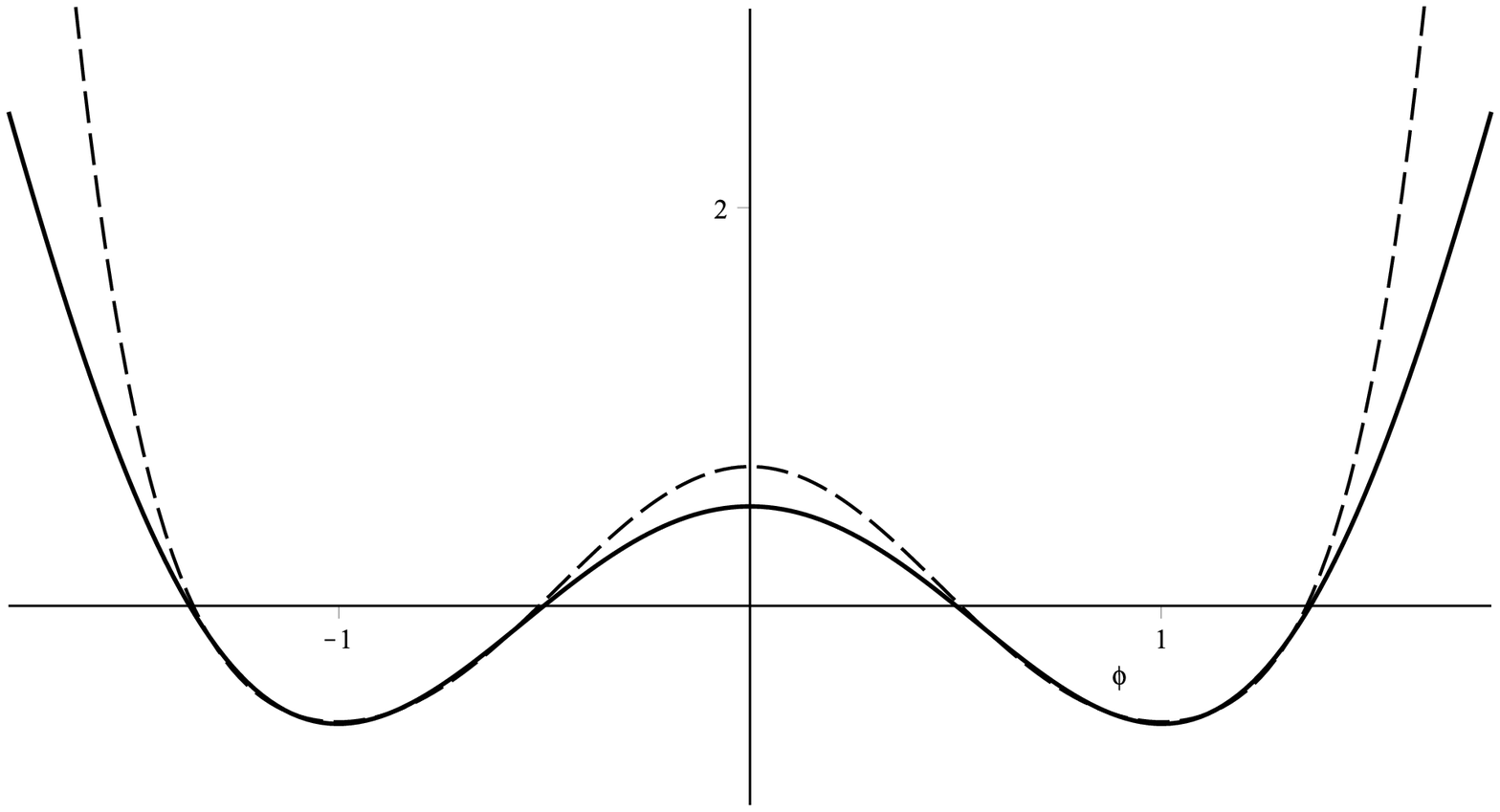}
\caption{The potential \ref{vaprox}, depicted for example 1 with $\epsilon=0$ (solid line), and with $\epsilon=-0.05$ and $n = 2$ (dashed line), taking $k^2=2$.}\label{fig:3}
\end{figure}
\begin{figure}[t]
\includegraphics[{height=3.0cm,width=8cm,angle=-00}]{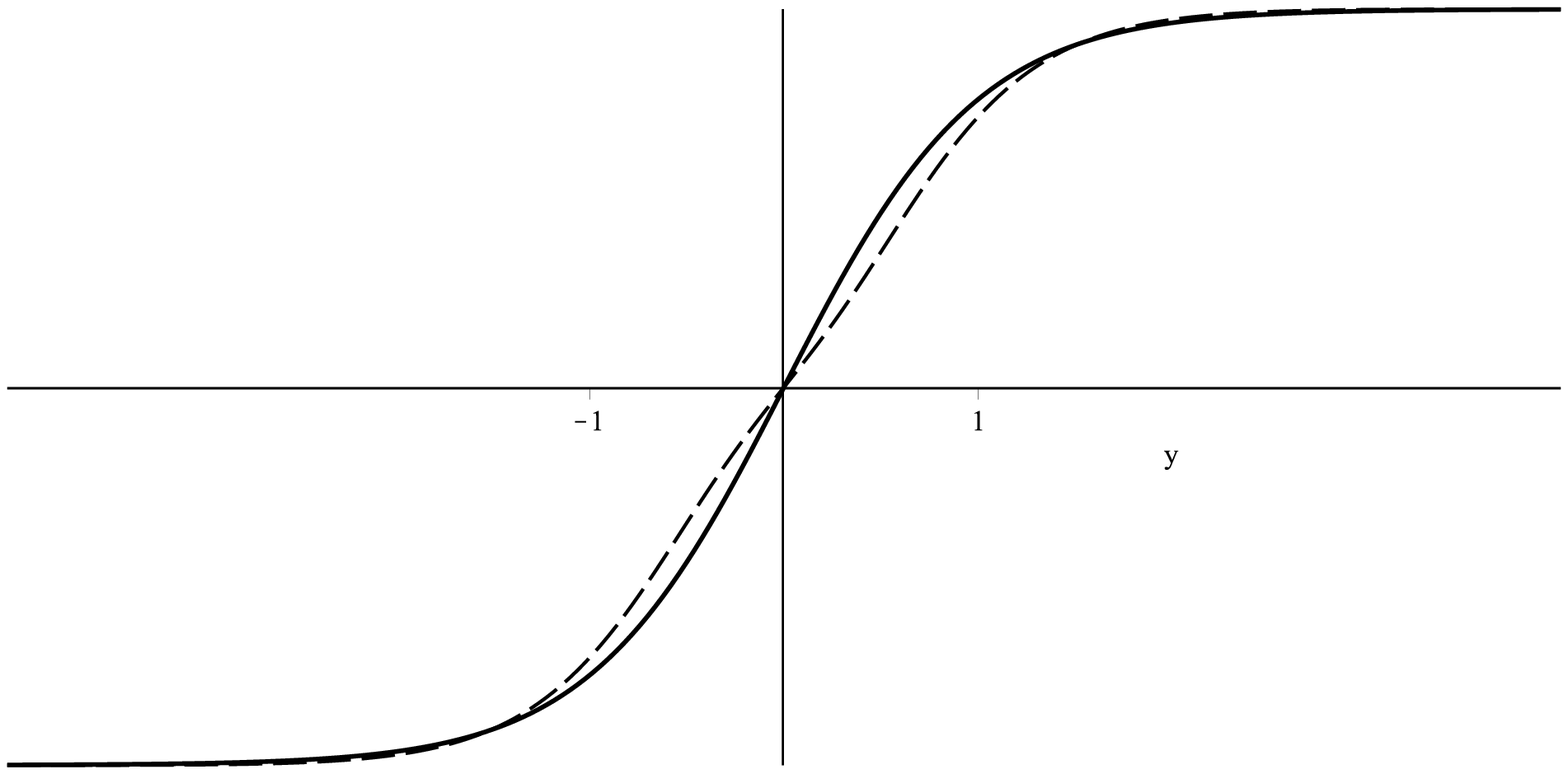}
\caption{The static solution \ref{phiaprox}, depicted for example 1 with $\epsilon=0$ (solid line), and with $\epsilon=-0.05$ and $n = 2$ (dashed line), taking $k^2=2$.}\label{fig:4}
\end{figure}

\begin{figure}[t]
\includegraphics[{height=3.0cm,width=8cm,angle=-00}]{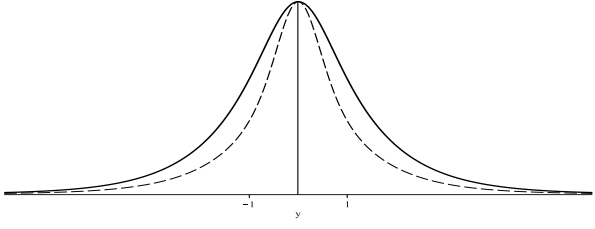}
\caption{The warp factor $\exp(2\,A(y))$ is obtained integrating numerically Eq.~\ref{Ayaprox}. It is depicted for example 1 with $\epsilon=0$ (solid line), and with $\epsilon=-0.05$ and $n = 2$ (dashed line), taking $k^2=2$.}\label{fig:5}
\end{figure}

\begin{figure}[t]
\includegraphics[{height=3.0cm,width=8cm,angle=-00}]{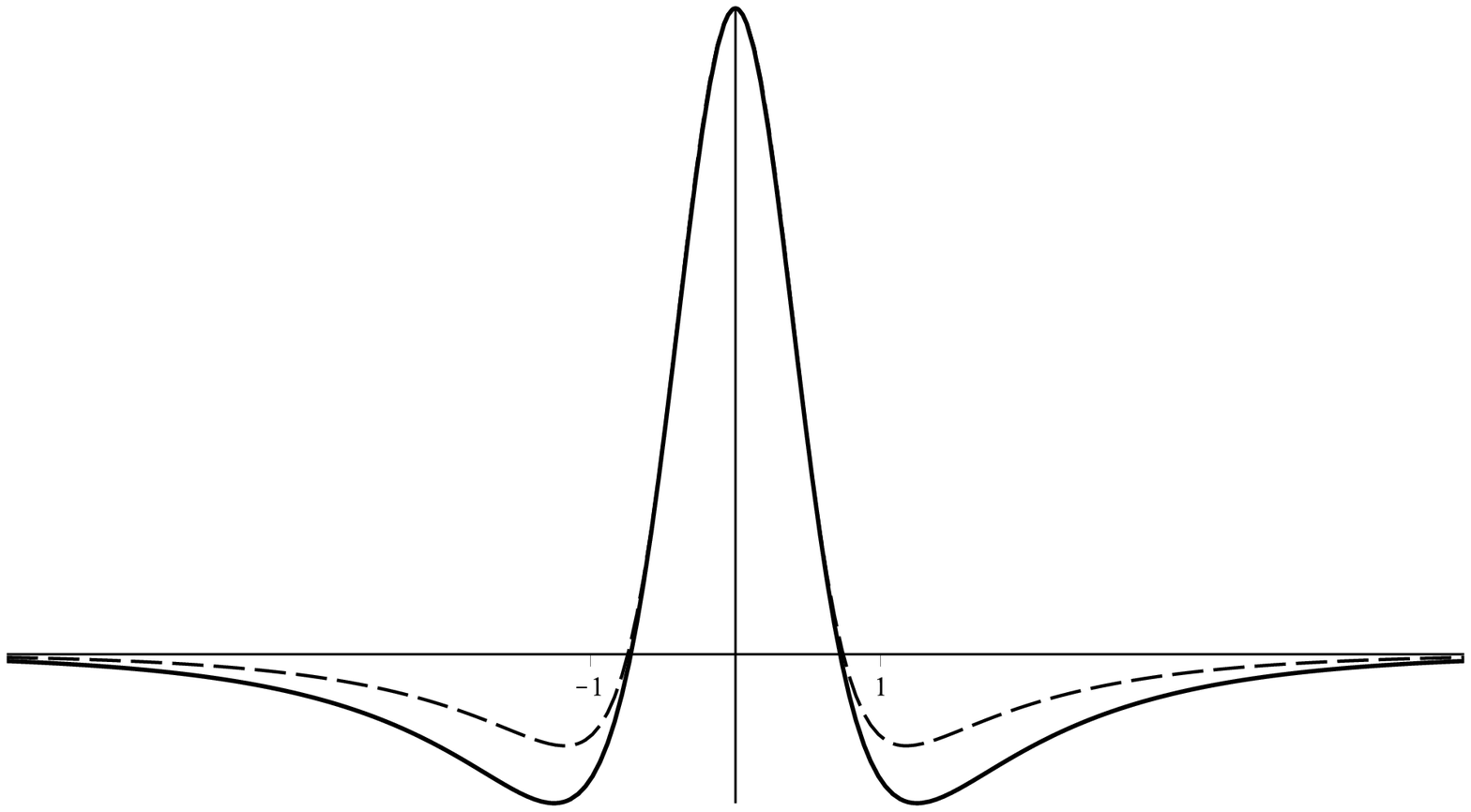}
\caption{The energy density, depicted for example 1 after integrating numerically Eq.~\ref{Ayaprox} and using Eq.~\ref{energy1}. We consider $\epsilon=0$ (solid line) and $\epsilon=-0.05$, with $n = 2$ (dashed line), taking $k^2=2$.}\label{fig:6}
\end{figure}

Moreover, from Eq.~\ref{eq:superphi} we have
\begin{equation}
\phi_y=\frac12W_\phi\left(1+\frac43\epsilon\,R_0^{n-1}\right)\,,
\end{equation}
and so
\begin{equation}
2\frac{d\phi}{W_\phi}\left(1-\frac43\epsilon\,R_0^{n-1}\right)=dy\,,
\end{equation}
or better
\begin{equation}\label{tildey}
2\int \frac{d\phi}{W_\phi}=y+\frac83\epsilon \int^{\phi_0} \frac{R_0^{n-1}}{W_\phi} d\phi=\tilde{y}\,.
\end{equation}
The case of $\epsilon=0$ gives
\begin{equation}
2 \int \frac{d\phi}{W_\phi}=y\,\to \phi=\phi_0(y)\,,
\end{equation}
where $\phi_0(y)$ is the solution for $\epsilon=0$, the unperturbed solution. Then, for $\epsilon\neq 0$, from \ref{tildey} we have
\begin{equation}
2\int\! \frac{d\phi}{W_\phi}=\tilde{y}\,\to \phi=\phi_0(\tilde{y})=\phi_0\left(y\!+\!\frac83\epsilon\! \int^{\phi_0} \frac{R_0^{n-1}}{W_\phi}
d\phi \right ).
\end{equation}
Expanding up to first-order in $\epsilon$ we obtain
\begin{equation}
\phi(y)=\phi_0(y)+\frac83\epsilon \frac{d\phi_0}{dy} \int^{\phi_0} \frac{R_0^{n-1}}{W_\phi} d\phi \,,
\end{equation}
or
\begin{equation}\label{phiaprox}
\phi(y)=\phi_0(y)+\epsilon \,\phi_\epsilon(y)\,,
\end{equation}
with
\begin{equation}
\phi_\epsilon(y)=\frac43\frac{d\phi_0}{dy} \int R_0^{n-1}(\phi_0)\,dy\,.
\end{equation}
Now, we use \ref{eq:Ay} to write
\begin{equation}\label{Ayaprox}
A_y=-\frac13W(\phi_0)-\frac13\epsilon\, Q(\phi_0)-\frac13 \epsilon\, \phi_\epsilon(y) \frac{d}{dy}(W(\phi_0))\,,
\end{equation}
where
\begin{equation}
Q=\frac13 W R_0+20n(n-1)\left(\frac{W_{\phi\phi}}{15}-\frac{W}{9}\right)W_\phi^2 R_0^{n-2}\,.
\end{equation}


Finally, the energy density $T_{00}$ can be written as
\begin{equation}\label{energy1}
\rho(y)=-e^{2A(y)}{\cal L}(y)\,,
\end{equation}
with ${\cal L}(y)$ given by
\be\label{deny}
{\cal L}(y)=-\frac12 \phi_0^{\prime 2} - V(\phi_0) - \epsilon \left(\phi_0 + \frac{d}{dy}V(\phi_0)\right).
\en
These are the general results, which we now illustrate with two distinct examples.

The first example is described by $W(\phi)=2\phi-2\phi^3/3$ and $n=2$. The potential \ref{vaprox} is illustrated in Fig. \ref{fig:3}. The static solution is given by \ref{phiaprox}, where $\phi_0=\tanh(y)$ and
\bea
\phi_\epsilon(y)&=&\frac{16}{729}{\rm sech}^2(y)\Big(141\tanh(y)-52\tanh^3(y)\nonumber\\
&&+3\tanh^5(y)-60\,y\Big)\,.
\ena
This solution is depicted in Fig.~\ref{fig:4}. The warp factor, which is obtained after integrating numerically Eq.~\ref{Ayaprox}, is displayed in Fig.~\ref{fig:5}. From this figure we see that the warp factor contributes to further localize the extra dimension, as we noted before in the example depicted in Fig.~\ref{fig:1}. Finally we have plotted the corresponding energy density in Fig.~\ref{fig:6}.

\begin{figure}[t]
\includegraphics[{height=3.0cm,width=8cm,angle=-00}]{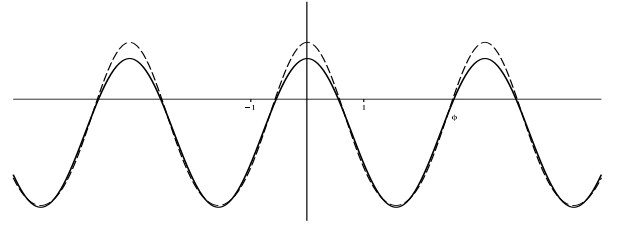}
\caption{The potential \ref{vaprox}, depicted for example 2, with $a=b=1$. We consider $\epsilon=0$ (solid line)  and $\epsilon=-0.05$, with $n = 2$ (dashed line), taking $k^2=2$.}\label{fig:7}
\end{figure}
\begin{figure}[t]
\includegraphics[{height=3.0cm,width=8cm,angle=-00}]{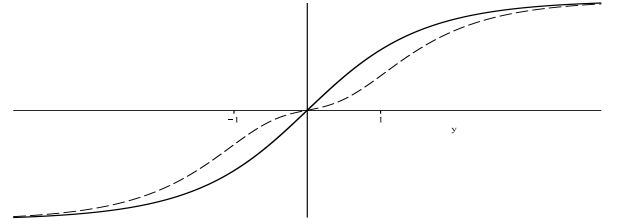}
\caption{The static solution \ref{phiaprox}, depicted for example 2 with $a=b=1$. We consider $\epsilon=0$ (solid line)  and
$\epsilon=-0.05$, with $n = 2$ (dashed line), taking $k^2=2$.}\label{fig:8}
\end{figure}

\begin{figure}[t]
\includegraphics[{height=3.0cm,width=8cm,angle=-00}]{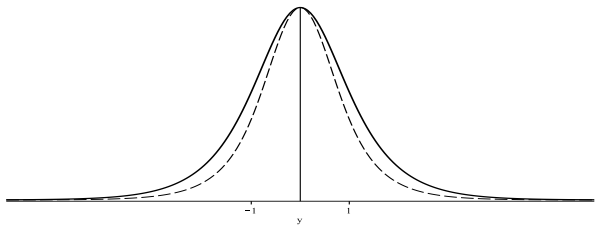}
\caption{The warp factor $\exp(2\,A(y))$ is depicted for example 2 with $a=b=1$, after integrating numerically Eq.~\ref{Ayaprox}. We consider
$\epsilon=0$ (solid line) and $\epsilon=-0.05$, with $n = 2$ (dashed line), taking $k^2=2$.}\label{fig:9}
\end{figure}

\begin{figure}[t]
\includegraphics[{height=3.0cm,width=8cm,angle=-00}]{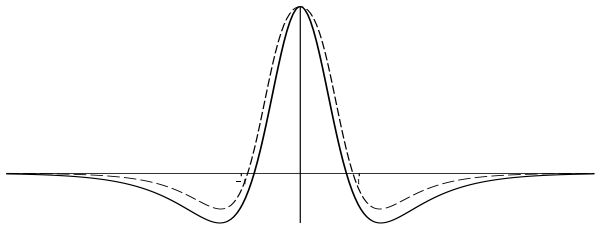}
\caption{The energy density, depicted for example 2 with $a=b=1$, after integrating numerically Eq.~\ref{Ayaprox} and using Eq.~\ref{energy1}. We consider $\epsilon=0$ (solid line)  and $\epsilon=-0.05$, with $n = 2$ (dashed line), taking $\kappa^2=2$.}\label{fig:10}
\end{figure}

The second example is described by $W(\phi)=2a^2 b\sin(\phi/a)$ and $n=2$. The potential \ref{vaprox} is illustrated in Fig.~\ref{fig:7}. The static solution around $V(\phi=0)$ is given by \ref{phiaprox}, where $\phi_0=a\arcsin(\tanh(y))$ and
\be
\phi_\epsilon(y)=\frac{16}{27}a^3b^2{\rm sech}(by)\left(\left(3+5a^2\right)\tanh(by)-5a^2b\,y\right)\,.
\en
which is depicted in Fig.~\ref{fig:8}. Again, the warp factor is obtained after integrating numerically Eq.~\ref{Ayaprox}, and it is displayed in Fig.~\ref{fig:9}. Finally, in Fig.~\ref{fig:10} we depict the respective energy density. Once again, we note that the warp factor depicted in Fig.~\ref{fig:9} shows the tendency to localize the extra dimension, as we move away from the standard scenario.

\section{Comments and conclusions}
\label{sec:end}

In this work we have investigated braneworld models in an $AdS_5$ warped geometry with a single extra spatial dimension of infinite extent. We implemented the study under the Palatini approach, in which both the metric and connection are treated as independent degrees of freedom. As the Palatini formulation for GR turns out to be equivalent to the standard metric approach (where the connection is taken a priori to be compatible with the metric), we had to replace the Einstein-Hilbert Lagrangian, $R$, by a nonlinear $f(R)$ Lagrangian, as the simplest extension of GR. The source action, which is responsible for the generation of the brane, was taken as that of a single scalar field in analogy with the standard thick brane scenario of GR.

We worked out the equations of motion for arbitrary $f(R)$ Lagrangian, which are second-order differential equations, and proved their consistency. A first-order framework to solve the equations of motion has been obtained by generalizing the approach used in the GR case. We then investigated a simple (and exact) example of a Palatini brane and showed that the warp factor engenders an interesting effect, the tendency to localize the extra dimension due to the nonlinear corrections.

We also studied another example of brane in which a small parameter is introduced to control the departures from the dynamics of GR. The small parameter was used to carry out a perturbative investigation. Although the perturbative approach cannot be used to probe the model in full detail, it has been useful to show the robustness of the brane solutions against Palatini $f(R)$ perturbations in the dynamics. As in the exact model studied in Sec. \ref{sec:example}, in the two perturbative examples that we investigated in Sec. \ref{sec:papproach}, we also noted that the warp factor tends to localize the extra dimension. We then conjecture that the thick braneworld scenario developed in the Palatini approach contributes to localize the extra dimension, an effect which is rather appealing and deserves further attention (see \cite{Gu} for a different approach on this problem). In this context, it would be interesting to study further extensions of GR including additional curvature invariants, as suggested by the quantization of fields in curved space-times \cite{quantization}, and in other braneworld scenarios. These investigations are currently underway.

\section*{Acknowledgments}

D.B., L.L., and R.M. would like to thank CNPq for financial support. G.J.O. is supported by a Ramon y Cajal contract, the Spanish grants FIS2014-57387-C3-1-P and FIS2011-29813-C02-02 from MINECO, the grants i-LINK0780 and i-COOPB20105 of the Spanish Research Council (CSIC) and the Consolider Program CPANPHY-1205388. D. R.-G. is funded by the Funda\c{c}\~ao para a Ci\^encia e a Tecnologia (FCT) postdoctoral fellowship No.~SFRH/BPD/102958/2014, the FCT research grant UID/FIS/04434/2013, and the NSFC (Chinese agency) grant No.~11450110403. The authors also acknowledge funding support of CNPq project No. 301137/2014-5.


\end{document}